\documentclass[a4paper,11pt]{article}
\usepackage{jheppub} 
\usepackage{array}
\usepackage{tikz}
\usepackage[compat=1.1.0]{tikz-feynman} 
\usepackage{amssymb}
\usepackage{xcolor}
\usepackage{slashed,phonetic}
\usepackage{hyperref}
\usepackage{standalone}
\usepackage{multirow}
\usepackage{appendix}
\usepackage{pgfplots}
\usepackage{pgfplotstable}
\pgfplotsset{compat=1.18}
\usepackage{multicol}
\usepackage{comment}

\newcommand{\SU}{\mathrm{SU}}

\newcommand{\SO}{\mathrm{SO}}
\newcommand{\U}{\mathrm{U}}

\title{\boldmath Leptogenesis in five-dimensional asymptotic Grand Unification models}

\author[a]{Timoth\'e Alezraa,}
\affiliation[a]{Universit\'e Claude Bernard Lyon 1, CNRS/IN2P3, IP2I UMR 5822,  4 rue Enrico Fermi, F-69100 Villeurbanne, France}

\author[b,c]{Giacomo Cacciapaglia,}
\affiliation[b]{Laboratoire de Physique Th\'eorique et Hautes \'Energies (LPTHE), UMR 7589, Sorbonne Universit\'e \& CNRS, 4 place Jussieu, 75252 Paris Cedex 05, France}
\affiliation[c]{Quantum Theory Center (QTC) \& D-IAS, Southern Denmark Univ., Campusvej 55, 5230 Odense M, Denmark}

\author[a,d]{Aldo Deandrea}
\affiliation[d]{Department of Physics, University of Johannesburg,
PO Box 524, Auckland Park 2006, South Africa.}

\emailAdd{timothe.alezraa@ens-paris-saclay.fr}

\abstract{Asymptotic grand unification models can be constructed in five dimensions compactified on an orbifold. We demonstrate that the parameter space of such models admit solutions that naturally achieve the baryon asymmetry via leptogenesis, these solutions indicating that the scale of the extra dimensions is much higher than the TeV scale.}

\begin{document}
\maketitle
\flushbottom

\section{Introduction}
\label{sec:intr}
An intriguing indication of new physics is the possibility of unification of gauge couplings, which may occur at high energies when considering the renormalization group evolution of these couplings in specific models. The approximate occurrence of this phenomenon in the standard model (SM) has inspired the development of Grand Unified Theories (GUTs) \cite{Georgi:1974sy,Fritzsch:1974nn}. The traditional paradigm of quantitative unification implies that the SM gauge couplings must be equal (up to group theory factors) at a specific energy scale, assuming that above that scale a new and larger gauge symmetry applies. The classic GUTs are based on $\SU(5)$ and $\SO(10)$, implying various degrees of unification for the Yukawa couplings. The unified gauge group must, therefore, be broken by a Higgs mechanism at a high scale, below which the gauge couplings of the SM are differentiated via their running. Achieving the correct numerical values at low energies always requires the presence of some light new particles (often due to supersymmetry \cite{Dimopoulos:1981yj}), while unavoidable proton decay pushes the unification scale above $\sim 10^{16}$~GeV \cite{Bajc:2002bv}. 
Another interesting source of new physics is the emergence of low-scale extra dimensions \cite{Antoniadis:1990ew}, primarily supported by string theory constructions. Extra dimensions change the running of couplings, hence offering new avenues for constructing models of quantitative unification \cite{Dienes:1998vh,Ghilencea:1998st,Hall:2002ea}. One of the most compelling ideas in this context was developed within gauge-Higgs unification models \cite{Hosotani:1983xw,Hatanaka:1998yp,Haba:2004qf}, allowing to protect the Higgs mass from large loop corrections. The two mechanism can also be combined in grand gauge-Higgs unification models \cite{Hosotani:2015hoa}.

The paradigm of asymptotic unification (aGUT) in five dimensions (5D) \cite{Cacciapaglia:2020qky} uses both ideas of unification and extra dimensions, without the requirement for precise coupling unification at a specific energy scale. Instead, the couplings tend asymptotically towards the same value at high energies. This is obtained by means of a non-trivial ultraviolet (UV) fixed point \cite{bajc_asymptotically_2016}. 5D models with a single extra spatial dimension may feature this property \cite{Gies:2003ic,Morris:2004mg}, which stems from the power-law running of the gauge couplings \cite{Dienes:2002bg}. The extra dimension is compactified on an interval $S^1/\mathbb{Z}_2\times\mathbb{Z}_2'$, where the bulk can be either flat or warped, as the high-energy behaviour remains consistent in both cases. Models of this type  exhibit a unified gauge symmetry in the bulk, leading to unified behaviour at high energies (small distances). The bulk gauge group is broken by boundary conditions, and the SM fields are identified with the zero modes of the bulk fields. This implies a distinct arrangement of matter and gauge fields within multiplets, differing from the standard multiplets of traditional GUT schemes with important phenomenological consequences, such as the absence of proton decay. A first toy model in the electroweak sector was proposed in \cite{Abdalgabar:2017cjw}, based on a bulk $\SU(3)$ symmetry. The first aGUT based on $\SU(5)$ was proposed in \cite{Cacciapaglia:2020qky,Cacciapaglia:2022nwt}, however, it was not possible to find a fixed point for all Yukawa couplings in the bulk. Further studies allowed to identify more realistic scenarios \cite{Khojali:2022gcq,Cacciapaglia:2023ghp,Cacciapaglia:2023kyz,Cacciapaglia:2024duu,Cacciapaglia:2025bxs}, whose limited number is explained by the strong requirements imposed by the aGUT model building. The only new scale in 5D aGUT models stems from the inverse radius of the extra dimension, which controls the Kaluza-Klein (KK) mass of the bulk resonances. In realistic aGUTs this scale can vary over many orders of magnitude, depending on the details of the model, typically from a few TeV  up to the more usual GUT unification scales. This property suggests that aGUT models may be tested both at future colliders and in different aspects such as baryogenesis, flavour and cosmology.

In standard GUT models, baryogenesis is closely related to leptogenesis \cite{Davidson:2008bu}. In this work, we shall examine the corresponding realisation of leptogenesis within aGUT models. As a general motivation, leptogenesis offers a simple and elegant explanation for the cosmological matter-antimatter asymmetry. The link between the baryon asymmetry and the lepton sector has important consequences for the neutrino sector. Leptogenesis is driven by the CP-violating interactions of the lightest heavy Majorana neutrinos, imposing stringent constraints on the masses of both light and heavy neutrinos. In our scenario, the Majorana mass is localised on the boundaries of the extra dimensions (fixed points of the orbifold), while all SM fields propagate in the extra dimension. Leptogenesis in extra dimensions has already been considered in models where only neutrinos propagate in the bulk, which is either flat \cite{Pilaftsis:1999jk,Abada:2006yd} or warped \cite{Medina:2006hi,Gherghetta:2007au}.

In the following, we explore the neutrino sector and the constraints induced by leptogenesis in a simple template aGUT model based on $\SU(5)$. This choice is motivated by the fact that realistic models always contain an $\SU(5)$ subgroup, hence they share the same qualitative features. In Section~\ref{sec:numass} we recap the main properties of the $\SU(5)$ aGUT, with focus on the neutrino sector, before discussing leptogenesis in Section~\ref{sec:leptog}. How to extend such results to more realistic cases is briefly addressed in Section~\ref{sec:realaGUT}.

\section{Neutrino masses in a template SU(5) aGUT}
\label{sec:numass}

The $\SU(5)$ aGUT of \cite{Cacciapaglia:2020qky,Cacciapaglia:2022nwt} fails to be a complete model for the lack of UV fixed points within the Yukawa sector. Nevertheless, it contains all the necessary ingredients of realistic models, of which it is a subset. Hence, we will use it as a simple and familiar template for the study of aGUT leptogenesis. Besides recapping the general structure of the model and the $\SU(5)$ multiplet embedding of the SM fermions, we will describe in more detail the neutrino sector, which was only briefly sketched in \cite{Cacciapaglia:2020qky}. Importantly, a see-saw mechanism can be embedded in the model thanks to a localised Majorana mass for the bulk right-handed neutrinos, whose treatment contains subtleties in order to ensure the correct decoupling limit.


\subsection{SU(5) aGUT model}

In this section, we briefly recap the field content of the $\SU(5)$ aGUT template, with focus on the embedding of the SM fields within multiplets, their parities under $\mathbb{Z}_2 \times \mathbb{Z}'_2$, and their charges. Note, in particular, that the matter multiplets cannot be arranged in the same way as the standard $\SU(5)$ GUT, due to the fact that the parities (i.e. the boundary conditions on the multiplet components) break $\SU(5)$. Hence, the minimal embedding requires the presence of bulk $\bf 5$-- $\bf \overline{5}$ and $\bf 10$--$\bf \overline{10}$ pairs, with different boundary conditions. Together with components that correspond to the SM fields (containing chiral massless zero modes, and massive KK modes), the bulk fields contain ``Indalo'' states, which do not feature zero modes at all.
The minimal field content for one generation, therefore, reads
\begin{equation}\label{eq:matter} \begin{split}
 \psi_{1_{L/R}} = N_{L/R}\,, \quad
    \psi_{5_{L/R}} = \begin{pmatrix} d \\ L^c
\end{pmatrix}_{L/R} \, , \quad
    \psi_{\overline{5}_{L/R}} = \begin{pmatrix} D^c \\ l
\end{pmatrix}_{L/R}  \, , & \\
 \psi_{10_{L/R}} = \frac{1}{\sqrt{2}}\begin{pmatrix} U^c & q \\ & E^c
\end{pmatrix}_{L/R} \,, \quad
\psi_{\overline{10}_{L/R}} = \frac{1}{\sqrt{2}}\begin{pmatrix} u & Q^c \\ & e
\end{pmatrix}_{L/R} \, , &
\end{split} \end{equation}
where $\psi$ indicates 5D bulk fermions, which have both a left-handed ($L$) and a right-handed ($R$) part (with opposite parities). The components are labelled according to their SM quantum numbers, with $q$ ($Q$) for the quark doublet, $u$ ($U$) for the up singlet, $d$ ($D$) for the down singlet, $l$ ($L$) for the lepton doublet, $e$ ($E$) for the charged lepton singlet, and $N$ for the right-handed neutrino. The lower case indicates the components containing (chiral) SM zero modes, while the upper case indicates the Indalo partners (with the exception of the right-handed neutrino $N$). Finally, the superscript ${}^c$ stands for charge conjugate.

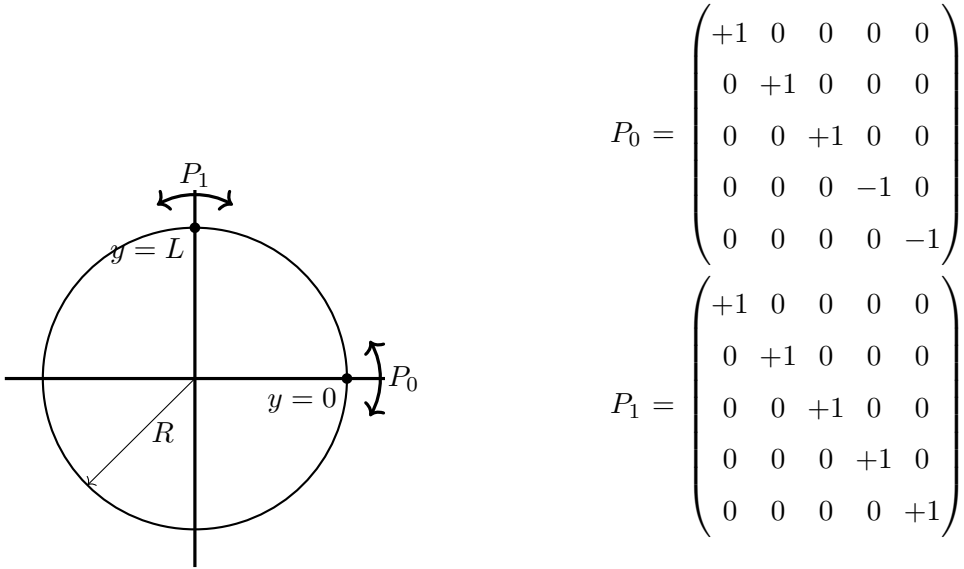
\begin{figure}[ht]
\centering
\begin{tabular}{*{2}{>{\centering\arraybackslash}b{\dimexpr0.5\linewidth-2\tabcolsep\relax}}}
\begin{tikzpicture}
  \draw[very thick] (-2.5,0) -- (2.5,0);
  \draw[very thick] (0,-2.5) -- (0,2.5);
  \draw[thick] (0,0) circle [radius=2cm];
  \draw(2,0) node[below left]{$y=0$};
  \draw(2.4,0) node[right]{$P_0$};
  \fill (2,0) circle[radius=2pt];
  \draw[very thick,<->] (2.3,0.5) arc  (30:-30:1cm) ;
  \draw(0,2) node[below left]{$y=L$};
  \draw(0,2.4) node[above]{$P_1$};
  \fill (0,2) circle[radius=2pt];
  \draw[very thick,<->] (0.5,2.3) arc  (90-30:90+30:1cm) ;
  \draw[->] (0,0) -- (-1.414,-1.414) node[midway,right] {$R$} ;
\end{tikzpicture}
    &
\renewcommand{\arraystretch}{1.3}

\begin{eqnarray*}
P_0 &=& \begin{pmatrix} +1&0&0&0&0\\ 0&+1&0&0&0\\ 0&0&+1&0&0\\ 0&0&0&-1&0\\0&0&0&0&-1
\end{pmatrix} \\
P_1 &=& \begin{pmatrix} +1&0&0&0&0\\ 0&+1&0&0&0\\ 0&0&+1&0&0\\ 0&0&0&+1&0\\0&0&0&0&+1
\end{pmatrix}
\end{eqnarray*}
\end{tabular}
\caption{\label{fig:1} Schematic representation of the fifth dimension orbifold, and explicit expression of the two parity matrices.}
\end{figure}

The matter content above stems from the parities needed to break the bulk $\SU(5)$ down to the SM gauge symmetry $\SU(3)_c\times\SU(2)_L\times\U(1)_Y$. The $\SU(5)$ gauge theory is defined on a $\mathcal{M} = \mathbb{R}^4 \times S^1/(\mathbb{Z}_2 \times \mathbb{Z}'_2) =\mathcal{\Tilde{M}} \times S^1/(\mathbb{Z}_2 \times \mathbb{Z}'_2)  $ orbifold. The two parities act around the fixed points $0,\pi/2$ and they are defined in terms of two $\SU(5)$ matrices $P_0,P_1$, so the fundamental domain of our theory is $\left[0,L\right] = \left[0,\pi R /2\right]$. A schematic illustration of the extra dimension can be found in Fig.~\ref{fig:1}, where we also indicate the explicit expression for the two parity matrices. It is clear that the $\SU(5)$ breaking is done by the first parity $P_0$, while $P_1$ preserves the bulk gauge symmetry.

\par For the gauge fields $A^a_M$, where $M = (\mu, y)$ is the 5D Lorentz index, this implies the following relations:
\begin{equation}
    (P_0) \Rightarrow \left\{ \begin{array}{l} A^a_{\mu}(x,-y)  = P_0 A^a_{\mu}(x,y) P_0^{\dagger}\,,
\\
    A^a_y(x,-y) = - P_0 A^a_y(x,y) P_0^{\dagger} \,;
\end{array} \right.
\end{equation}
\begin{equation}
    (P_1) \Rightarrow \left\{ \begin{array}{l} A^a_{\mu}(x,\pi R - y) = P_1 A^a_{\mu}(x, y) P_1^{\dagger}\,,
    \\
    A^a_y(x,\pi R - y) = - P_1 A^a_y(x, y) P_1^{\dagger} \,;
\end{array} \right.
\end{equation}
where $A_y$ is the polarization along the fifth compact dimension, and the fields are periodic over $y \to y + 2 \pi R$. The radius $R$ defines the mass scale of the KK modes of each field. With the choice of parities in Fig.~\ref{fig:1}, the bulk $\SU(5)$ is broken to the SM group on the $y=0$ boundary. We also recall that parity-odd fields respect Dirichlet boundary conditions (i.e. vanishing of the field), while parity-even ones respect Neumann boundary conditions (i.e. vanishing of the derivative of the field). Thus, all the gauge-scalar modes are ``eaten" by the massive vector KK modes. The modes with parities $(+,+)$ correspond to the usual SM gauge fields and their KK modes, while the off-diagonal gauge fields do not have a zero mode and, therefore, correspond to Indalo states. Their properties are listed at the bottom of Table~\ref{tab:1}.

\par The Higgs boson is embedded in  $\bf 5$ bulk scalar field, with a QCD-triplet scalar $H$, 
\begin{equation}
    \phi_5 = \begin{pmatrix} H \\ \phi_h \end{pmatrix} \,,
\end{equation}
with parities:
\begin{equation}\label{eqn:2-5}
    (P_0) \Rightarrow \phi_5(x,-y) = - P_0 \phi_5(x,y) \, ,
\end{equation}
\begin{equation}\label{eqn:2-6}
    (P_1) \Rightarrow \phi_5(x,\pi R - y) = + P_1 \phi_5(x, y) \, .
\end{equation}
Hence the QCD triplet $H$ belongs to the Indalo sector, it has no zero mode and the usual problem of doublet-triplet splitting is absent.


For the matter multiplets, the parities of the 5D 4-component spinors are chosen as follows:
\begin{equation}\label{eqn:2-8}
    (P_0) \Rightarrow \left\{ \begin{array}{l} \psi_1(x,-y) = - \gamma_5 \psi_1(x,y) \,,
    \\\psi_5(x,-y) = + P_0 \gamma_5 \psi_5(x,y) \,,
    \\
    \psi_{\overline{5}}(x,-y) = + P_0^{\dagger} \gamma_5 \psi_{\overline{5}}(x,y) \,, \\
    \psi_{10}(x,-y) = + P_0 \gamma_5 \psi_{10}(x,y) P_0^{T} \, , 
    \\
    \psi_{\overline{10}}(x,-y) = + P_0^{\dagger} \gamma_5 \psi_{\overline{10}}(x,y) P_0^{*} \,; 
\end{array} \right.
\end{equation}
\begin{equation}\label{eqn:2-9}
    (P_1) \Rightarrow \left\{ \begin{array}{l} \psi_1(x,\pi R - y) = - \gamma_5 \psi_1(x, y) \,,
    \\\psi_5(x,\pi R - y) = + P_1 \gamma_5 \psi_5(x, y) \,,
    \\
    \psi_{\overline{5}}(x,\pi R - y) = - P_1^{\dagger} \gamma_5 \psi_{\overline{5}}(x, y) \,, \\
    \psi_{10}(x,\pi R - y) = - P_1 \gamma_5 \psi_{10}(x, y)  P_1^{T} \,,
    \\
    \psi_{\overline{10}}(x,\pi R - y) = + P_1^{\dagger} \gamma_5 \psi_{\overline{10}}(x, y)  P_1^{*} \, .
\end{array} \right.
\end{equation}
In the above equations $\gamma_5$ is the Dirac matrix in the (diagonal) Weyl representation. It encodes the different sign for the L and R components under the four-dimensional reduction. The parity choice above leads to the matter fields introduced in Eq.~\eqref{eq:matter}.

The most general gauge-invariant Lagrangian in the bulk is given by the following:
\begin{eqnarray}
 \mathcal{L}_{SU(5)} &= & -\frac{1}{4}F^{(a)}_{M N}{F^{(a)}}^{M N} - 
 \frac{1}{2\xi}(\partial_{\mu}A^{\mu} - \xi \partial_{5}A_y)^2 + i \overline{\psi_{5}}\slashed{D}\psi_5 + i \overline{\psi_{\overline{5}}}\slashed{D}\psi_{\overline{5}} + i \overline{\psi_{10}}\slashed{D}\psi_{10}  \notag \\
&+& i \overline{\psi_{\overline{10}}}\slashed{D}\psi_{\overline{10}} - \left( \sqrt{2} Y_{e}\,  \overline{\psi_{\overline{5}}} \psi_{\overline{10}}\phi_5^* + \sqrt{2} Y_{d}\,  \overline{\psi_{5}} \psi_{10}\phi_5^* + \frac{1}{2} Y_{u}\,  \epsilon_5\ \overline{\psi_{\overline{10}}} \psi_{10}\phi_5 + \mbox{h.c.}\right)  
 \notag \\
 &+& |D_M \phi_5|^2 - V (\phi_5) + i \overline{\psi_1}\slashed{\partial}\psi_{1} - \left( Y_\nu\, \overline{\psi_1} \psi_{\bar{5}} \phi_5 + \mbox{h.c.} \right)\,, \label{Model-Lagrangian}
\end{eqnarray}
where, besides the gauge interactions and the gauge fixing term, we have included all the Yukawa couplings allowed in the model. One can immediately see that the theory permits one coupling per SM fermion, so that no unification occurs in the Yukawa sector. It has also been shown in \cite{Cacciapaglia:2020qky} that one can define a unique conserved global charge, which matches the baryon number for the SM fermions. This feature is enough to forbid dangerous proton decay operators. In Table~\ref{tab:1} we summarise the properties of all the fields, including the baryon and lepton number assignments (knowing that lepton number is violated by some of the interactions within the Yukawa sector). The states in boxes contain the chiral SM fermions as their zero modes.

\begin{table}[ht]
\begin{center}
\begin{tabular}{ |p{1.6cm}||p{1cm}||p{1cm}|p{1cm}|p{1.5cm}|p{2.5cm}| }
 \hline
 Multiplets    & Fields  & L & B &$(\mathbb{Z}_2, \mathbb{Z}'_2)$ & SM \\
 \hline
 $\psi_{\overline{5}}$ & $D_R^c$ & 1/2 & 1/6 & $(-,+)$ & \multirow{1}{*}{$({\bf 3}, {\bf 1}, -1/3)$} \\
  &  \fbox{$e_L$}  & 1 & 0 & $(-,-)$ & $\in$ \multirow{1}{*}{$({\bf 1}, {\bf 2}, -1/2)$}\\
  &  \fbox{$\nu_L$}  & 1 & 0 & $(-,-)$ & $\in$ \multirow{1}{*}{$({\bf 1}, {\bf 2}, -1/2)$}\\
 \hline
 $\psi_{5}$ & \fbox{$d_R$} &  0 & 1/3 & $(+,+)$ &$\in$  \multirow{1}{*}{$({\bf 3}, {\bf 1}, -1/3)$} \\
 &  $E_L^c$  & -1/2 & 1/2 & $(+,-)$ &$\in$  \multirow{1}{*}{$({\bf 1}, {\bf 2}, -1/2)$}\\
 &  $\mathcal{N}_L^c$  & -1/2 & 1/2 & $(+,-)$ & $\in$  \multirow{1}{*}{$({\bf 1}, {\bf 2}, -1/2)$}\\
 \hline
  $\psi_{10}$ & $U_R^c$ &  1/2 & 1/6 & $(-,+)$ & $\in$  \multirow{1}{*}{$({\bf 3}, {\bf 1}, 2/3)$}\\
  & $E_R^c$ & -1/2 & 1/2 & $(-,+)$ &   \multirow{1}{*}{$({\bf 1}, {\bf 1}, -1)$}\\
  & \fbox{$u_L$} &  0 & 1/3 & $(-,-)$ &  $\in$  \multirow{1}{*}{$({\bf 3}, {\bf 2}, 1/6)$}\\
  & \fbox{$d_L$} &  0 & 1/3 & $(-,-)$ &  $\in$  \multirow{1}{*}{$({\bf 3}, {\bf 2}, 1/6)$}\\
  \hline
  $\psi_{\overline{10}}$ & \fbox{$u_R$} & 0 & 1/3 & $(+,+)$ &  $\in$  \multirow{1}{*}{$({\bf 3}, {\bf 1}, 2/3)$}\\
  & \fbox{$e_R$}  & 1 & 0 & $(+,+)$ & \multirow{1}{*}{$({\bf 1}, {\bf 1}, -1)$}\\
  & $U_L^c$ & 1/2 & 1/6 & $(+,-)$ &  $\in$  \multirow{1}{*}{$({\bf 3}, {\bf 1}, 2/3)$} \\
  & $D_L^c$ & 1/2 & 1/6 & $(+,-)$ &  $\in$  \multirow{1}{*}{$({\bf 3}, {\bf 1}, -1/3)$} \\
 \hline
 $\psi_1$ & $N$ & $1$ & $0$ & $(-,-)$ &    \multirow{1}{*}{$({\bf 1}, {\bf 1},0)$} \\ 
 \hline
 $\phi_5$ & $H$ & 1/2 & -1/6 & $(-,+)$ &  \multirow{1}{*}{$({\bf 3}, {\bf 1}, -1/3)$}\\
 & \fbox{$\phi^+$} & 0 & 0 & $(+,+)$ &  $\in$  \multirow{1}{*}{$({\bf 1}, {\bf 2}, 1/2)$}\\
 & \fbox{$\phi_0$} & 0 & 0 & $(+,+)$ & $\in$  \multirow{1}{*}{$({\bf 1}, {\bf 2}, 1/2)$}\\
 \hline
  $A_X$ & $X_\mu$ & 1/2 & -1/6 & $(-,+)$ & $\in$  \multirow{1}{*}{$({\bf 3}, {\bf 2}, -5/6)$}\\
 & $Y_\mu$ & 1/2 & -1/6 &  $(-,+)$ & $\in$  \multirow{1}{*}{$({\bf 3}, {\bf 2}, +5/6)$}\\
 \hline
\end{tabular} \end{center}
\caption{\label{tab:1} Baryon and lepton numbers for the components of the $\SU(5)$ multiplets. We also indicate their parity under orbifold and the embedding into the SM groups. All Indalo states have baryon numbers half of the SM ones, hence they cannot decay into purely SM states. The lightest Indalo may, therefore, play the role of dark matter \cite{Cacciapaglia:2020qky}.}
\end{table}

As a final remark, we recall that the expansion in KK modes derives from the propagation of the field in the extra dimension, allowing to decompose the 5D fields in a tower of four-dimensional fields with specific wave functions that depend on the extra coordinate:
\begin{equation}
    \phi(x^\mu, y) = \sum_n \ f_n (y) \ \varphi_n (x^\mu)\,. 
\end{equation}
The form of the wave functions depend on the geometry of the space, i.e. whether it is warped or flat. In the following, for simplicity, we will focus on a flat extra dimension, where the warped case is a straightforward extension. What determines the wave functions, and the corresponding KK masses, are the parities:
\begin{equation} \label{eq:wavefunctions}
\begin{split}
    (+,+)& : f_n(y) = \cos(n\frac{\pi}{L}y) \,, \quad m_n = \frac{n}{R}\,,\\
    (-,-)& : f_n(y) = \sin((n+1)\frac{\pi}{L}y) \,, \quad m_n = \frac{n+1}{R}\,,\\
    (+,-)& : f_n(y) = \cos[(n+1/2)\frac{\pi}{L}y ] \,, \quad m_n = \frac{n+1/2}{R}\,,\\
    (-,+)& : f_n(y) = \sin[(n+1/2)\frac{\pi}{L}y] \,, \quad m_n = \frac{n+1/2}{R}\,,
\end{split}
\end{equation}
where $n  \in  \mathbb{N}$. We see that only parities $(+,+)$ allow to contain a zero mode, hence fermionic zero modes are necessarily chiral.

\subsection{Neutrino see-saw with a brane-localised Majorana mass}

The model, as presented so far, only allows for Dirac neutrino masses, given by the Yukawa coupling $Y_\nu$, as the bulk interaction decomposes as:
\begin{equation}
    Y_\nu\ \overline{\psi_1} \psi_{\overline 5} \phi_5 = Y_\nu\ \left( \overline{N_R} \phi_h l_L +  \overline{N_L} \phi_h l_R + \overline{N_R} H B_L^c + \overline{N_L} H B_R^c\right)\,,
\end{equation}
so that the zero mode in $N_R$ couples via the Higgs to the zero mode in $l_L$.
This is mainly due to the fact that no Majorana masses can be added in the bulk, as Majorana spinors do not exist in 5D. Nevertheless, we are left with the freedom to add Majorana masses for the singlet $N$ on the boundaries, where the localised Lagrangian needs only to respect Poincar{\'e} invariance in four dimensions.

For concreteness, we added the Majorana mass on the $y=L=\pi R/2$ fixed point, where the $\SU(5)$ symmetry is not broken. To ensure a consistent derivation of the boundary conditions that assure the decoupling limit \cite{Angelescu:2019viv,Leng:2020ofk}, we also include appropriate bilinear boundary terms (BBT). They are required when deriving the equations of motion in the bulk and on the boundaries, see \cite{Nortier:2020xms} for a pedagogical derivation. Finally, the model is completed by the following two localised pieces of action:     
\begin{align}
     S_{\rm Maj}  = & - \int_\mathcal{M} d^4x\; dy \;\delta(y-L)\ \frac{L}{2} \ \left( \overline{\psi_1^c}_R M \psi_{1R} +  \overline{\psi_1^c}_L M' \psi_{1L}\right)\,, \\
     S_{\rm BBT} = & - \sum_{f} \int_\mathcal{M} d^4x\; dy\; [\delta(y) - \delta(y-L)] \frac{\mu_f(y)}{2} (\overline{\psi_f} \psi_f +h.c)\,;
\end{align}
where $f$ stands for all the fermions in the bulk.
As it is not consistent to have both non-vanishing $M$ and $M'$, we set $M'=0$ consistently with the parity assignments on $\psi_1$ (according to which $\psi_{1L} = 0$ on the boundaries). In the following, we focus on the neutrino sector, and we well use $N = \psi_1$ to denote the singlet field.

The boundary conditions can be obtained by the vanishing of the surface variation of the action:
\begin{equation}\label{EL1}
[\delta \overline{N}_L N_R  - \delta \overline{N}_R N_L]^{y=L}_{y=0} = M L \ \delta \overline{N}_R N_R^c \mid_{y=L} \pm [\mu_N (y) (\delta \overline{N}_L N_R  - \delta \overline{N}_R N_L)]^{y=L}_{y=0}\,.
\end{equation}
For $\delta N \neq 0$, this yields 
\begin{equation}
\begin{split}
    & ( 1+\mu_N(0))\ N_R \mid_{0}= 0\,, \\
    & ( 1-\mu_N(L))\ N_R \mid_{L}=0 \,, \\
    & ( 1-\mu_N(0))\ N_L \mid_{0}=0\,, \\
    & ( 1+\mu_N(L))\ N_L - ML \ N_R^c \mid_{L}=0\,.
\end{split}
\end{equation}
To match the parities of the bulk field $\psi_1$, we choose $\mu_N (0) = -1$ and $\mu_N(L)=1$, so that the non-trivial boundary conditions become:
\begin{equation}
    N_L \mid_{0}=0\,, \qquad 2\ N_L - ML \ N_R \mid_{L}=0\,.
\end{equation}
As the boundary conditions mix the two chiralities of the bulk field, we can expand the two components in terms of the same tower of (Majorana) states as follows:
\begin{equation}
    \begin{split}
        N_L (x^\mu, y) &= \frac{1}{\sqrt{L}} \sum_n f_{L,n}(y)\ N_n(x^\mu)\,, \\
        N_R (x^\mu, y) &= \frac{1}{\sqrt{L}} \sum_n f_{R,n}(y)\ N^c_n(x^\mu)\,, 
    \end{split}
\end{equation}
where $N_n$ are four dimensional Majorana fermions with KK frequency $k_n$, which is determined by an interplay of equations of motion and boundary conditions. The wave functions, $f_{L/R,n}$, stem from the bulk equation of motion:
\begin{equation}
    i\Gamma^M \partial_M N + Y_\nu \phi_h l =0\,,
\end{equation}
involving $\nu_L$ via the Higgs vacuum expectation value. For simplicity, we will neglect the effect of $Y_\nu$ in the KK expansion and consider it as a small perturbation. Hence, the approximate wave functions are given by
\begin{equation} 
    f_{L/R,n} = A_{L/R,n} \cos (k_n y) + B_{L/R,n} \sin (k_n y)\,,
\end{equation}
where no periodicity condition is needed. After imposing the parities and boundary conditions, and canonically normalising the four dimensional fields, we obtain:
\begin{equation} \label{eq:wfN}
    f_{L,n}(y) = - \frac{1}{\sqrt{2}} \sin (k_n y)\,, \quad f_{R,n} (y) = \frac{1}{\sqrt{2}} \cos (k_n y)\,,
\end{equation}
with
\begin{equation}
    (M_n \equiv )\ k_n = \frac{1}{L} \left( \arctan (ML) + n \pi \right)\,, \quad n \in \mathbb{Z}\,.
\end{equation}
In the absence of the bulk Yukawa $Y_\nu$, the mass of each mode is given by $M_n \equiv k_n$. Corrections to the masses and, more importantly, a mixing with the KK modes in $\nu_L$ will be generated via the bulk Higgs coupling.

\begin{figure}[tbh]
\centering 
\begin{tabular}{cc}
\begin{tikzpicture}
\begin{feynman} 
\vertex (a) {\( \nu_{L}^0 \)};
\vertex [right=of a,dot,label=below:\(Y_{0i}\)] (b){};
\vertex [above left=of b] (h1) {\( \phi_h \)} ;
\vertex [right=of b,crossed dot,label=\( M_{i}\)](c){};
\vertex [right=of c,dot,label=below:\(Y_{0i}\)](d){};
\vertex [right=of d] (f2) {\(\nu_{L}^0 \)};
\vertex [above right=of d] (h2) {\( \phi_h \)};
\diagram* {
(a) -- [fermion] (b) -- [scalar] (h1),
(b) -- [fermion, edge label'=\(N_i\)](c),
(c) -- [anti fermion, edge label'=\(N_i\)](d),
(d) -- [anti fermion] (f2),
(d) -- [scalar] (h2),
};
\end{feynman}\end{tikzpicture} & 
\begin{tikzpicture}
\begin{feynman}
\vertex (a) {\( \nu_{L}^0 \)};
\vertex [right=of a,dot,label=below:\(Y_{0i}\)] (b){};
\vertex [above left=of b] (h1) {\( \phi_h \)} ;
\vertex [right=of b,dot,label=below:\( Y_{ji}\)](c){};
\vertex [above=of c](h11){\( \phi_h\)};
\vertex [right=of c,dot,label=below:\(Y_{jk}\)](d){};
\vertex [above=of d](h12){\( \phi_h\)};
\vertex [dot][right=of d,label=below:\(Y_{0k}\)] (e) {}; 
\vertex [right=of e] (f2) {\(\nu_{L}^0 \)};
\vertex [above right=of e] (h2) {\( \phi_h \)};
\diagram* {
(a) -- [fermion] (b) -- [scalar] (h1),
(b) -- [plain, edge label'=\(N_i\)](c),
(c) -- [plain, edge label'=\textcolor{red}{\(\nu_L^j\)}](d),
(c) -- [scalar](h11),
(d) -- [scalar](h12),
(d) -- [plain, edge label'=\(N_k\)] (e),
(e) -- [anti fermion] (f2),
(e) -- [scalar] (h2),
};
\end{feynman}
\end{tikzpicture}
\end{tabular}
\caption{\label{fig:numass} Leading (left) and typical sub-leading (right) diagrams leading to a Majorana mass for the zero-mode left-handed neutrinos. In the second diagram, a Majorana mass insertion is understood either on $N_i$ or $N_k$.}
\end{figure}
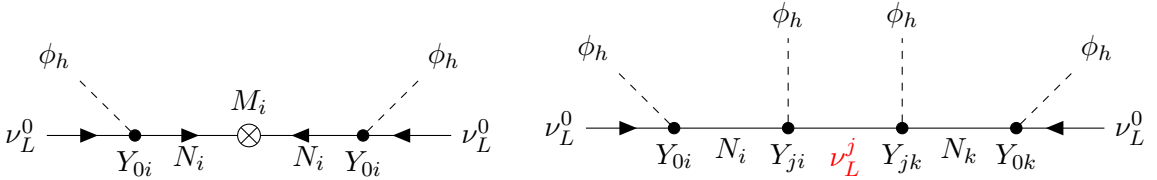

The (Majorana) mass of the SM neutrino can be computed in an expansion for small $Y_\nu$ by integrating out the heavy singlets $N_n$. In Fig.~\ref{fig:numass} we show the leading diagram to the left, and a typical subleading one to the right, where $Y_{ij}$ are the four dimensional couplings stemming from the bulk Yukawa as follows:
\begin{equation}
    S_{Y_\nu} = \int_{\mathbb{R}^4 }d^4x \sum_{k,j}
    \underbrace{\int_0^L dy\; \frac{Y_\nu v}{L^{3/2}}\ f_{R,j}(y) f_{l,k}(y)}_{Y_{kj} }\;
    N_j(x^\mu)\nu_L^k (x^\mu)\,,
\end{equation}
where $f_{l,k} = \frac{1}{\sqrt{2L}} \cos (k \pi y/L)$ is the normalised wave function for the neutrino field and $v$ is the SM Higgs vacuum expectation value (which we assume to be independent on $y$).
As all KK modes have masses of the same order, Fig.~\ref{fig:numass} clearly shows that the leading order dominates as long as $v \ll 1/L$,  as $1/L$ is at least of the order of a few TeV \cite{Cacciapaglia:2020qky}.

Hence, we can neglect the effect of the KK modes of $\nu_L$, and only include its zero mode. This greatly simplifies the form of the mass matrix, which now reads, in the basis $(\nu_L^0, N_0, \dots N_n \dots)$:
\begin{equation}
    \mathbf{M} = \lim_{n\to \infty} \begin{pmatrix}
0 & a_0 & a_1 & a_{-1} & \cdots & a_n  & a_{-n}\\
a_0 & M_0 & 0 & 0 & \cdots & 0 & 0\\
a_1 & 0 & M_1 & 0 & \cdots & 0  & 0\\
a_{-1} & 0 & 0 & M_{-1} & \dots & 0 & 0\\
\vdots & \vdots & \vdots & \vdots & \ddots & \vdots & \vdots\\
a_n &0 & 0 & 0 & \cdots & M_n & 0 \\
a_{-n} &0 & 0 & 0 & \cdots & 0& M_{-n}  
\end{pmatrix}\,,
\end{equation}
with 
\begin{equation}
    a_i = Y_{0i} v\,.
\end{equation}
The smallest eigenvalue of the above matrix gives an estimate of the SM neutrino (Majorana) mass:
\begin{equation} \label{eq:mnu}
        m_\nu \equiv \lambda_{min} \approx \sum_{i \in \mathbb{Z}} \frac{a_i^2 }{M_i}   
        = \left( \frac{Y_\nu v}{\sqrt{2}} \right)^2 \cot(\arctan(ML)) 
        =  \frac{1}{M} \frac{(Y_\nu v)^2}{2 L} \,.
\end{equation}
The series was computed using the residue theorem, $ \sum_\mathbb{Z}f(n) = \sum_{\text{poles}} \text{Res}(\pi \cot{\pi z}f(z))  $, hence the cotangent function, consistent  with the arctan  term in the mass.
From the above equation, we can see that a see-saw of type-I is in place thanks to the interplay between the bulk Yukawa (we recall that $Y_\nu$ scales like $\sqrt{L}$) and the localised Majorana mass $M$. Also, the one-flavour analysis we presented here can be easily extended to a multi-flavour scenario.

\section{KK mode leptogenesis}
\label{sec:leptog}




The measured baryon asymmetry of the universe is usually expressed in terms of the density of baryons minus the density of antibaryons, $n_B$, normalised either by the entropy density or by the density of photons. Its value can been extracted from the Planck cosmological observations of $\Omega_b \, h^2 =0.0224 \pm 0.0002$ \cite{Planck:2018vyg} to give \cite{ParticleDataGroup:2024cfk}:
\begin{equation}
    \eta_B \equiv \frac{n_B}{n_\gamma} = (6.12\pm 0.04) \cdot 10^{-10} \,.
\end{equation}
Assuming that the asymmetry is generated above the electroweak phase transition, the value of $\eta_B$ can be related to the density of $B-L$ (baryon minus lepton charge) density in a comoving volume containing one photon $N_{B-L}$ as \cite{Buchmuller:2004nz}
\begin{equation}
    \eta_B \sim 10^{-2}\; N_{B-L}\,.
\end{equation}
Hence, our aim in this section is to estimate the $B-L$ asymmetry generated in the minimal $\SU(5)$ aGUT via the decays of the right-handed neutrinos, according to the standard leptogenesys scenario \cite{Buchmuller:2004nz,Davidson:2008bu}. As all KK modes contribute, we will follow the formalism from Ref.~\cite{Eisele:2007ws}, to which we refer the reader for more details.

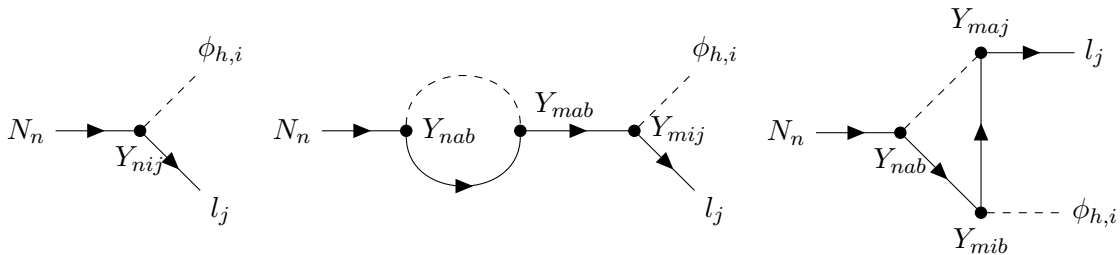
\begin{figure}[h!]
    \centering
      \begin{tikzpicture}[baseline=(current bounding box.center)]
    \begin{feynman}
\vertex (a) {\( N_n \)};
\vertex [right=of a,dot,label=below:\( Y_{nij}\)] (b){};
\vertex [above right=of b] (c) {\( \phi_{h,i} \)} ;
\vertex [below right=of b](d){\( l_{j}\)};
\diagram* {
(a) -- [fermion] (b) -- [scalar] (c),
(b) -- [fermion](d),
};
\end{feynman}
  \end{tikzpicture}
  \begin{tikzpicture}[baseline=(current bounding box.center)]
\begin{feynman}
\vertex (a) {\( N_n \)};
\vertex [right=of a,dot,label=right:\( Y_{nab} \)] (b){};
\vertex [right=of b,dot,label=above right:\( Y_{mab} \)] (c) {} ;
\vertex [right=of c,dot,label=right:\( Y_{mij} \)](d){};
\vertex [above right =of d](h1){\( \phi_{h,i}\)};
\vertex [below right =of d](f1){\( l_j\)};
\diagram* {
(a) -- [fermion] (b) -- [scalar,half left] (c),
(b) -- [fermion,half right](c),
(c) -- [fermion](d),
(d) -- [scalar] (h1),
(d) -- [fermion] (f1),
};
\end{feynman}
  \end{tikzpicture}
  \begin{tikzpicture}[baseline=(current bounding box.center)]
    \begin{feynman}
\vertex (a) {\( N_n \)};
\vertex [right=of a,dot,label=below:\(  Y_{nab} \)] (b){};
\vertex [above right=of b,dot,label=above:\(Y_{maj}  \)] (c) {} ;
\vertex [below right=of b,dot,label=below:\( Y_{mib}\)](d){};
\vertex [ right=of c](f1){\( l_{j}\)};
\vertex [ right=of d](h1){\( \phi_{h,i} \)};
\diagram* {
(a) -- [fermion] (b) -- [scalar] (c),
(b) -- [fermion](d),
(d) -- [fermion](c),
(c) -- [fermion](f1),
(d) -- [scalar](h1),
};
\end{feynman}
  \end{tikzpicture}
    \caption{ \label{fig:asymdiags} $N_n$ decays at tree level and one loop. Particles created in loops can be Indalos or SM-like.}
\end{figure}

One possible source of asymmetry is in the decays of the KK modes of the singlet $N$, generated by a possible CP violating phase in the Yukawa coupling $Y_\nu$. We can define an asymmetry parameter for each KK mode as follows:
\begin{equation} \label{eq:epsilonn}
    \epsilon_n = \frac{ \sum_{X}  \Gamma(N_{n}\rightarrow X) - \Gamma(N_{n}\rightarrow \bar{X})
    }{ \sum_{X} \Gamma(N_{n}\rightarrow X) + \Gamma(N_{n}\rightarrow \bar{X})}\,,
\end{equation}
where $X$ contains all possible final states with lepton number $L=1$. In the $\SU(5)$ model, we have two decay modes:
\begin{equation}
    X = \phi_{h,i} l_j\,, \;\; H_i D^c_{R,j}\,,
\end{equation}
where $i,j$ label the kinematically allowed KK modes. Hence, each decay will be ruled by the effective Yukawa couplings
\begin{equation}
    Y_{nij} = \int_0^L\ dy\ \frac{Y_\nu}{L^{3/2}}\ f_{n}(y) f_{i}(y) f_{j}(y)\,, 
\end{equation}
where $f$ are the appropriate wave functions for $N$, Eq.~\eqref{eq:wfN}, and the other fields, Eq.~\eqref{eq:wavefunctions}.
At zero temperature, the asymmetries $\epsilon_n$ vanish at tree level, so they must stem from loop effects.
Fig.~\ref{fig:asymdiags} shows typical diagrams for the $\phi_h l$ mode.

We also assume that leptogenesis starts at an early time after the end of inflation and reheating, where the universe is in a hot thermal bath. The singlets $N_n$ are also in equilibrium thanks to the Yukawa coupling to the Higgs and the lepton doublet fields. Hence, the evolution of the relevant densities can be traced by Boltzmann differential equations. For our purposes, we need to track the density of each KK singlet $N_n$ and of the $B-L$ charge density. We work under the assumption that the dominant processes change the $N$--densities by one unit. Hence the Boltzmann equations only depend on the densities of each mode, while the densities of other states remaining in thermal equilibrium are factored in the coefficients. In terms of densities per comoving volume containing a single photon, the Boltzmann equations read
\begin{eqnarray} \label{eq boltzmann many N}
	\frac {dN_{N_n}}{dz} &=& - \frac {1}{\mathcal{H}z} 
	(\Gamma _{D,n} + \Gamma _{\Delta L =1,n}) \, 
	(N_{N_n}-N^\textrm{eq}_{N_n}) \, , \\
\label{eq boltzmann many B-L}
	\frac {dN_{B-L}}{dz} &=& 
	- \frac {1}{\mathcal{H}z} \sum _{n=1}^{n_\textrm{eff}(z)} \left[
	 \epsilon _n (\Gamma _{D,n} + \Gamma _{\Delta L =1,n}) \, 
	(N_{N_n}-N^\textrm{eq}_{N_n})\,
	+ \Gamma _{W,n} N_{B-L} \right] \, ,
	\end{eqnarray}
where $\mathcal{H}=\mathcal{H}(z)$ is the expansion rate and $z = L^{-1}/T$ the inverse temperature (which we normalise with the interval length $L$). The details of the model are contained in the coefficients of the left-hand-side terms. Besides the asymmetry $\epsilon_n$ defined in Eq.~\eqref{eq:epsilonn}, we have:
\begin{itemize}
    \item $\Gamma_{D,n}$ is the (total) decay width of the mode $N_n$,
    diluted by thermal effects as $\Gamma_{D,n}(z) = \Gamma \times \frac{K_1(z)}{K_2(z)}$. 
    In the Boltzmann equation, the widths always appear as ratios $\frac {\Gamma}{\mathcal{H}z}$. Hence, we can define 
    \begin{equation}
    D_n(z)=\frac {\Gamma_{D,n}}{\mathcal{H}z}= K_n\times z \frac{K_1(z)}{K_2(z)}\,,\quad \mbox{where} \quad K_n=\frac {\Gamma_{D,n}(z=+\infty)}{\mathcal{H}(z=1)}
    \end{equation}
    is called the washout parameter. It indicates if the decays happen far or close to equilibrium. The washout parameter can be computed by summing all the amplitudes of tree level decay channels $N_n\to  \phi_{h,i} l_j\,,  H_i D^c_{R,j}\, $ and the expansion rate, as in \cite{Buchmuller:2004nz}.
    \item $\Gamma_{\Delta L,n}$ is the rate of lepton number violating processes, dominated by $2\leftrightarrow 2$ processes. It has been shown that the term appearing in the Boltzmann equations is always proportional to $D_n$.
    \item $\Gamma_{W,n}$ is the washout rate, from inverse decays and $\Delta L =1,2$ processes. Here again, they are proportional to the $D_n$ term. 
\end{itemize}

\begin{figure}[tbh]
    \centering
    \pgfplotstableset{
    create on use/X/.style={create col/copy column from table={data-plots/x_axis}{0}},
    create on use/Y1/.style={create col/copy column from table={data-plots/mode0_profile}{0}},
    create on use/Y2/.style={create col/copy column from table={data-plots/mode40_profile}{0}},
    create on use/Z/.style={create col/copy column from table={data-plots/BLevolution_with63modes}{0}}
}
\pgfplotsset{xmin=1e-4, xmax =1e3, ymin=1e-20, ymax=1e1}
\begin{tikzpicture}
        \begin{loglogaxis}[legend style={nodes={scale=0.6, transform shape}},title=Integrated Boltzmann equations,
                            ylabel=Densities,
                            xlabel={$z= L^{-1}/T$} , no markers]
            \addplot table[x =X ,y=Y2]{data-plots/mode40_profile};
            \addlegendentry{\(N_{40}\)}
            \addplot table[x =X ,y=Y1]{data-plots/mode0_profile} ;
            \addlegendentry{\(N_{0}\)}
            \addplot[dashed] table[x =X ,y=Z]{data-plots/BLevolution_with63modes};
            \addlegendentry{\(N_{B-L}\)}
        \end{loglogaxis}   
\end{tikzpicture}
    \begin{tikzpicture}
        \begin{semilogyaxis}[title=Final asymmetry as a function of the number of KK modes,
                            ylabel=$N_{B-L}(t\rightarrow{\infty})$,
                            xlabel=$k$]
            \addplot[only marks, mark=x] table[x expr=\coordindex ,y index=0]{data-plots/final_asym_outlier.txt};
        \end{semilogyaxis}   

\end{tikzpicture}
    
    \caption{Numerical results for the benchmark $L^{-1}= 10^{8}\, \mbox{GeV} ,\ M= 10^{10}\, \text{GeV},\ Y_\nu=10^{-9}\,  \text{GeV}^{-1/2}$. Left plot:  Evolution of the densities as a function of $z$ for the 0th and 40th KK modes, and the $B-L$ evolution for 40 modes. Right plot: Evolution of the final $B-L$ density as a function of the included number of modes $k$.}
    \label{res1}
\end{figure}
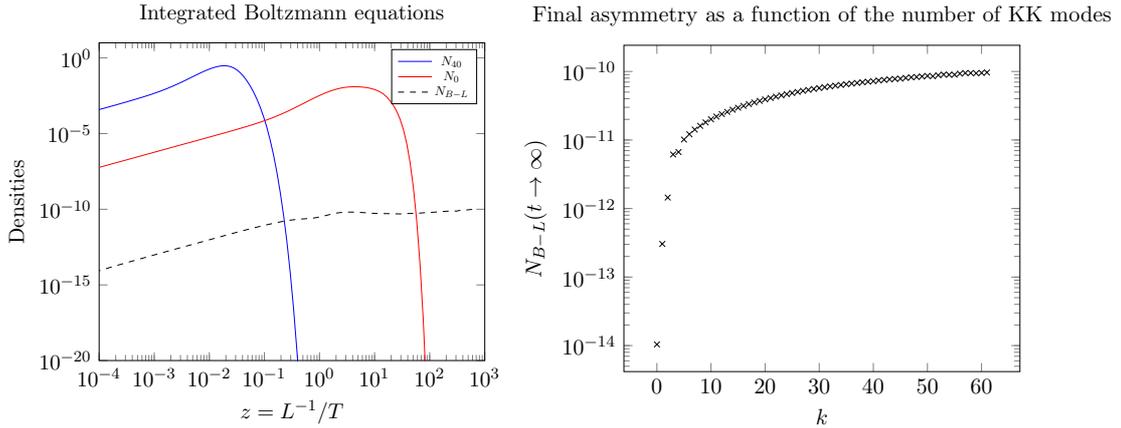

As we see, most of the dependency on the specifics of the model is contained in $D_n$, or equivalently in the washout parameters $K_n$. 
To obtain reliable solutions, we solve the Boltzmann equations numerically. Another model parameter is contained within the asymmetries $\epsilon_n$, which are sensitive to the CP violation in the neutrino Yukawa sector. Such parameters must satisfy a modified version of the Davidson-Ibarra bound \cite{Davidson:2008bu}, which, for our specific case where the singlet is propagating in the bulk, takes the form 
\begin{equation}\label{DI_bound}
    | \epsilon_n | \leq \frac{3}{16\pi}\frac{M_n}{v^2}m_{\nu}\,.
\end{equation}
In our numerical computations we will always take values saturating the bound.

The aGUT model under consideration has three free parameters entering the computation: the 5D Yukawa coupling $Y_\nu$, the length of the interval $L$ and the localised Majorana mass $M$. They are related to the value of the neutrino masses via Eq.~\eqref{eq:mnu}. In Fig.~\ref{res1} we show the evolution of the number densities for two sample KK modes, $N_0$ and $N_{40}$, and the evolution of the total $B-L$ number. In the plot, we used the following benchmark values:
\begin{equation}
    L^{-1}= 10^{8}\ \text{GeV}\,, \;\; M= 10^{10}\ \text{GeV}\,,\;\; Y_\nu=4\times 10^{-9}\  \text{GeV}^{-1/2}\,,
\end{equation}
which give $ m_\nu = 10^{-5}$~eV.
We see that the final $B-L$ asymmetry receives the latest contributions from the lightest modes, which decouple at a later time, i.e. at smaller temperatures $T$.
To test the dependence of the result on the truncation in the KK modes, we show in the right plot of  Fig.~\ref{res1} the final asymmetry value including $k$ modes in the computation. We see that the contribution is dominated by the first $10\div 20$ modes. For the numerical benchmark, the final result $N_{B-L} \approx 10^{10}$ falls about two orders of magnitude short of the experimental value.

\begin{figure}
    \centering
    \pgfplotstableset{
    create on use/X/.style={create col/copy column from table={data-plots/Y5_values}{0}},
    create on use/Y1/.style={create col/copy column from table={data-plots/final_asym_Y5}{0}}
}
\pgfplotsset{xmin=1e-11, xmax =1e-6, ymin=1e-19, ymax=1e1}
\begin{tikzpicture}
        \begin{loglogaxis}[grid,legend style={nodes={scale=0.6, transform shape}},title=Final Asymmetry as a function of $Y_\nu$,
                            ylabel=$N_{B-L}(t\to\infty)$,
                            xlabel={$Y_\nu$} , no markers]
            \addplot[] table[x =X ,y=Y1]{data-plots/mode0_profile} ;
        \end{loglogaxis}   
\end{tikzpicture}
    \caption{Final $B-L$ asymmetry as a function of $Y_\nu$, while keeping everything else fixed. We used the benchmark values $L^{-1}= 10^{8}\ \text{GeV} ,\;\; M= 10^{10}\ \text{GeV}$, including 10 modes.}
    \label{res3}
\end{figure}
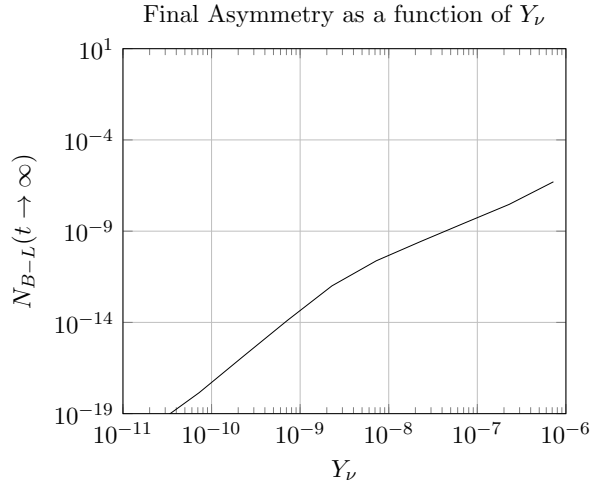

\begin{figure}[tbh]
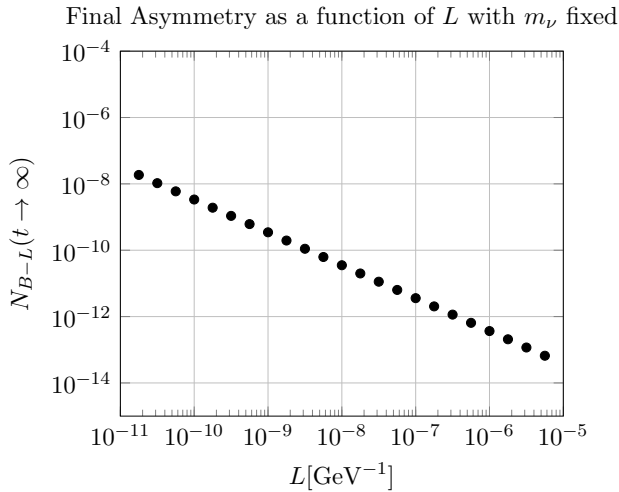

    \centering
    \includestandalone[scale=0.85]{data-plots/plot4}
    \caption{Final $B-L$ asymmetry as a function of $L$ while keeping fixed the neutrino mass. The result is obtained with the lightest $m_\nu= 10^{-5}$~eV  and $M= 10^{10}$~GeV, including 20 modes.}
    \label{res4}
\end{figure}

To understand the parameter dependence of this result, we first recall that for $M \gg L^{-1}$, the Majorana mass only enters in the determination of the neutrino mass \eqref{eq:mnu}, as the KK masses are determined purely by $L$. Instead, the computation for the final asymmetry depends on $Y_\nu$ and $L$ in two ways:
\begin{itemize}
    \item The asymmetries $\epsilon_n \propto m_\nu M_n$ enter the Boltzmann equation for $B-L$ \eqref{eq boltzmann many B-L}, hence increasing it leads to an increase in the final asymmetry.
    \item The rates proportional to $K_n \propto M_n Y_\nu^2$. Increasing them keeps the singlets in thermal equilibrium for longer via Eq.~\eqref{eq boltzmann many N}, hence reducing the final result.  
\end{itemize}
We first study the dependence on $Y_\nu$ by keeping all other parameters fixed (hence, $m_\nu \propto Y_\nu^2$). The result in Fig.~\ref{res3} shows how $N_{B-L}$ increases with $Y_\nu$, hence for the benchmark point we considered one could expect larger contributions to the asymmetry for heavier neutrinos, reaching realistic values for $Y_\nu \sim 10^{-7}\ \mbox{GeV}^{-1/2}$, which gives neutrino masses close to the scale from oscillations, $m_\nu \sim 10^{-2}$~eV.
If we fix the value of the neutrino masses, the dependence on the KK mass $L^{-1}$ is shown in Fig.~\ref{res4}, showing that the asymmetry finally increases with the KK mass, due mainly to the increase in the asymmetry parameters $\epsilon_n$. Nevertheless, the KK scale cannot be pushed to be too light in order to achieve the correct asymmetry.

\section{Extension to other aGUTs}
\label{sec:realaGUT}

Within the minimal $\SU(5)$ aGUT we have been focusing on so far, it is not possible to find UV fixed points for all Yukawa couplings that give mass to the SM fermionic zero modes \cite{Cacciapaglia:2020qky}. Although the model could still be of interest for phenomenology \cite{Cacciapaglia:2022nwt}, it does not provide a complete aGUT scenario. A thorough survey of possible aGUTs based on stable orbifolds \cite{Cacciapaglia:2023kyz,Cacciapaglia:2024duu,Cacciapaglia:2025bxs} revealed that there exists a single fully consistent model based on $\SU(6)$ in the bulk that leads to the SM directly at low energies.~\footnote{Other consistent models, based on $\SU(8)$ \cite{Cacciapaglia:2024duu}, $\SO(10)$ \cite{Khojali:2022gcq} and $E_6$ \cite{Cacciapaglia:2023ghp,Cacciapaglia:2025bxs}, feature an extended Pati-Salam model below the compactification scale.}

In this model, the orbifold breaks $\SU(6) \to \SU(3)\times\SU(2)\times\U(1)^2$, i.e. the SM gauge symmetry plus an additional $\U(1)$ factor. The latter can be broken either by Higgsing or by the localised anomalies \cite{Cacciapaglia:2024duu}. The left-handed leptons are embedded in a bulk $\bf 15$ with parities $(+,-)$, together with the quark doublet and other Indalo states. Instead, the Higgs is embedded in a scalar $\bf 15$ with parities $(-,+)$, together with other components that do not have zero modes. The relevant components and their parities are listed in Table~\ref{tab:2}.

\begin{table}[ht]
\begin{center}
\begin{tabular}{ |p{1.5cm}|p{1.5cm}||p{1.5cm}|p{1.5cm}||p{2.5cm}| }
 \hline
 \multicolumn{2}{|c||}{$\Psi_{\bf 15}$}   & \multicolumn{2}{c||}{$\Phi_{\bf 15}$}  & SM \\
 \hline
 $q$ & $(+,+)$ & $\phi_Q$ & $(-,-)$ & $({\bf 3}, {\bf 2}, 1/6)$ \\ \hline
 $l^c$ & $(-,-)$ & $\phi_h$ & $(+,+)$ & $({\bf 1}, {\bf 2}, 1/2)$ \\ \hline
 $E^c$ & $(+,-)$ & $\phi_{\bar E}$ & $(-,+)$ & $({\bf 1}, {\bf 1}, 1)$ \\ \hline
 $U^c$ & $(+,-)$ & $\phi_{\bar U}$ & $(-,+)$ & $({\bf \bar 3}, {\bf 1}, -2/2)$ \\ \hline
 $D$ & $(-,+)$ & $\phi_D \equiv H$ & $(+,-)$ & $({\bf 3}, {\bf 1}, -1/3)$ \\ \hline
\end{tabular} \end{center}
\caption{\label{tab:2} Components of the fields that couple to the singlets $N$ in the $\SU(6)$ aGUT model. We recall that for fermions $(+,+)$ implies a left-handed KK zero mode while $(-,-)$ a right handed one. For scalars, only $(+,+)$ leads to a zero mode.}
\end{table}

The singlets $N$ are introduced via a bulk singlet $\Psi_{\bf 1}$ with parities $(+,+)$ so that a zero mode appears with left-handed chirality. Like in the $\SU(5)$ model, the neutrino Yukawa coupling reads
\begin{equation}
    \mathcal{L}_{\rm Yuk} \supset Y_\nu \overline{\Psi_1} \Phi_{\bf 15}^\ast \Psi_{\bf 15} + \mbox{h.c.}
\end{equation}
The parity choice for the singlet $\Psi_1$ matches the fact that the left-handed lepton doublet is embedded in the bulk $\Psi_{\bf 15}$ with opposite chirality, $l^c$. 

Henceforth, the only difference with respect to the $\SU(5)$ model is that the singlets couple to a few more channels, which enter the computation of the decay width:
\begin{equation}
    X = l \phi_h^\ast\,, \;\; \bar{q} \Phi_Q\,, \;\; E \Phi_{\bar{E}}\,, \;\; U \phi_{\bar{U}}\,, \;\; \bar{D} H\,. 
\end{equation}
Roughly, this will lead to an enhancement by a factor of $5/2$ in the total width of the modes $N_n$, which does not affect significantly the results in the previous section.

\section{Conclusions}
\label{sec:concl}

Models of asymptotic grand unification offers an alternative and complementary realisation of the grand unification of gauge couplings in the SM. Models in five dimensions make use of the power-law running of the couplings to define UV fixed points in their renormalisation group evolutions. However, the strong requirements of having a non-trivial UV fixed point imply the existence of only a handful of feasible models.

In this work, we have demonstrated that aGUTs in 5D can provide a successful source of leptogenesis via a localised Majorana mass for the bulk neutrino singlets. We studied in detail the minimal $\SU(5)$ realisation, showing that similar results hold for the more realistic $\SU(6)$ aGUT model. Leptogenesis can provide sufficient baryon asymmetry at the price of raising the scale of the extra dimensions well above the TeV scale, hence out of reach for present and future colliders.

\acknowledgments
We thank Florian Nortier for his friendly support and for discussion and advice on a few technical points on extra-dimensional field theories with brane-localized terms.


\bibliographystyle{JHEP}
\bibliography{biblio.bib}

\end{document}